# FORMATION AND RUPTURE OF THE NANOSIZED METAL FILAMENT INSIDE OXIDE MATRIX

G. B. Stefanovich, M. J. Lee, B. S. Kang, S.-E. Ahn, K. H. Kim, C. B. Lee, C.J. Kim, and Y.S. Park

*Samsung Electronics, Samsung Advanced Institute of Technology, Gyeonggi-do 446-712, Republic of Korea*

The paper presents a model of the electrically actuated formation and rupture of the nanosized metal filament inside oxide matrix. The dielectric breakdown of the oxide with conditions of the adequate current compliance and subsequent capacitance discharge of the energy which have been stored in thin film oxide capacitor structure before breakdown results in sharp local temperature growth and, as a result, in fast local oxide reduction. The Soret state with metal segregation on the center of the high temperature region is established by temperature gradient-driven diffusion. The nanosized metal filament is quenched by fast temperature drop after capacitance discharge ending. At the next biasing, the local domain with high resistance and high electric field is created near cathode end of filament by metal electromigration due to electron wind induced by high density electron current. The part of metal filament is transformed in oxide by the subsequent fast electric field enhanced thermal oxidation.

The study was initiated by the experimental observation of the electrically actuated nonvolatile switching between two resistance states in thin films oxide structures [1]. To date this phenomenon is considered as promising candidate for development of the high density stackable nonvolatile memory. The nonvolatile switching usually can be divided on 2 different groups - polarity dependence switching and practically symmetrical unipolar switching. There is common understanding of the bipolar switching with memory. The majority of the researches show that bipolar switching is interface phenomenon in which the interface properties (interface transition layer resistance [2] or Shottky barrier height and form [3]) are modified by high field ionic transport. The mechanism of the unipolar switching with memory is far from understanding.

The most investigated oxide, at least in experimental sense, is NiO [4, 5, 6, 7, 8]. The experimental results show that first any polarity electrical biasing of the initial oxide structure with semiconductor type of the conductivity induces the growth of the thin filament (or filaments) with metal conductivity inside oxide matrix (forming). The second any polarity electrical biasing can ruptures metal filament that returns semiconductor properties to structure [4, 5, 6, 7, 8]. Further, transition between high resistance state (HRS) and low resistance state

(LRS) can be repeated many times.

The typical current-voltage characteristics (IVC) of the metal-oxide-metal (MOM) structure on base magnetron sputtering NiO with Pt top and bottom electrodes (TE and BE, respectively) which was using for modeling are shown in Fig.1. Also Fig. 1 demonstrates a cross section of the structure, giving geometry and notations.

The IVC of the initial oxide structure (Fig. 2) during first polarization measured at current controlled regime shows that forming can be classified as irreversible threshold switching with unstable section of the current controlled negative differential resistance (NDR). These features of the forming allow considering one as hard breakdown of the insulator oxide. Note that followed LRS-HRS transition can be observed only if adequate compliance current, $I_C$, have been used. The electrical biasing with high magnitude of the current compliance transfers structure in LRS state which can not be ruptured by next voltage input. The lowest level of the compliance current is defined by value of the breakdown threshold current.

There is not universal mechanism of thin films breakdown, but all researches allocate 2 process stages. At the first stage the sudden reduction of the insulator resistance driven by electronic or electrothermal positive feedback mechanism occurs. The section with NDR appears on IVC and the thin conductive path (cord) is formed between electrodes. During second stage of the breakdown the permanent conductive filament whose structure and chemical composition differs from native oxide is formed inside insulator [9]. Taking into attention this universal phenomenological behavior of the thin film insulators, the first breakdown stage is not so important for presented model. When any electronic or electrothermal instability has been initiated and, as a result, the conductive cord has been formed, the Joule heating increases temperature of this local region that can be resulted in local thermochemical modification of the oxide.

There are two approaches to the estimation of the energy dissipation region size. The universal thermodynamic consideration [10] shows that in system with initial uniform current distribution the trend of the current to collect in local domain is governed by the principal of least entropy production. Using approximation which have been developed in [10], we calculate that radius $a$ of cross-sectional area of the cylinder conductive cord, which have been formed after first breakdown stage, is 5nm.

Another approach is based on strong nonuniform distribution of the current in pre-breakdown state of the defect insulator [12]. The statistical model of the electric field enhancement by local geometric thinking of the oxide thickness assumes that the size of the high conductive path after first stage of the breakdown is the same as interface irregularities size – 5 nm [11]. Note also that the other high conductive defect in polycrystalline NiO is the grain boundaries which size have been measured as 5-10 nm [11, 13]. Consequently, the 5 nm as the dimension scale for *a* is a reasonable estimation.

Evidently there are two energy sources for Joule heating of the conductive domain. At first, it is necessary to taking into account the action of the direct current trough conductive cord. The density of the dissipated power can be calculated as $P_{DC} = I_c * V_c / v$, where $I_c$ is current compliance, $V_c$ is voltage which corresponds to $I_c$, and $v$ volume of the conductive domain. Obviously that $V_c \leq V_F$, where $V_F$ is forming voltage, because direct after first breakdown stage the structure has IVC with current-controlled NDR. Accepting a high current domain as cylindrical body with basis radius $a$ = 5 nm and oxide thickness as height $\delta$ = 50 nm we obtain $P_{DC} = 5*10^{13}$ W*cm$^{-3}$.

Before breakdown the structure is the capacitor with capacitance of *C* which is charged up to voltage of $V_F$. At the second breakdown stage, this energy is liberated by electrical discharge through conductive cord. The storage energy can be written as $E_C = [C*(V_F - V_C)^2]/2$ and one is equal to $10^{-13}$ J for analyzed sample. The capacitance discharge power density $P_C$ changes during energy liberation process but we will be allowed to assume that capacitance discharges occurs with constant rate at characteristic time $\tau_0 = C \cdot V_c/I_c = 10^{-9}$s and with this assumption the power density $P_C = E_C/(v*\tau_0) \approx 10^{16}$ W*cm$^{-3}$, therefore $P_C >> P_{DC}$. Let us note that $\tau \leq 10^{-9}$s and it is typical transient time of second breakdown stage for many thin insulator films [9, 12].

For estimations of the temperature space-time distributions we will assume, that the heat production is confined to a conductive cylinder with height $\delta$ and radius $a$. The temperature will be determined by oxide thermal conduction in both axial and radial direction, and spreading thermal resistance in electrodes. When *a* is enough small in comparison with other dimensions of structure (thickness of oxide, thickness and size of metal electrodes) we can assume that the temperature of electrodes and volume of the oxide far enough from the conductive path is equal ambient. Also, for thin conductive path, the cylinder lateral surface is much greater basis surface and we can assume that the radial heat flow will dominate.

The thermal time constant of the system $\tau_T$ can be estimated as $\tau_T = \delta^2/A$ [14], where $A = K_{NiO}/(c_{NiO} \gamma_{NiO})$ is NiO thermal diffusivity, where $K_{NiO}$ is thermal conductivity, $c_{NiO}$ specific heat, $\gamma_{NiO}$ density. The time constant $\tau_T$ is less then $10^{-10}$s and we will be allowed to assume that the steady state conditions are reached during capacitance discharge.

The steady-state solution of the heat equation in cylindrical coordinates with heating due to a line source of strength $Q_C = P_C/\delta$ along cylinder axis and with $T = 0$ at $r = d$, $dT/dr = 0$ at r = 0 is [14]

$$\Delta T_C = \frac{Q_C}{2\pi K_{NiO}} \ln\left(\frac{d}{r}\right), \qquad (1)$$

In practice $d$ will not be the sample or electrodes size and to obtain a realistic estimation we therefore replace $d$ with the oxide thickness $\delta$. Taking $K_{NiO} = 0.71$ W/(cm$^0$C) we obtain rise of temperature on the conductive cord boundary $(r = a)$ $\Delta T_C \approx 4000^0$C that is more then oxide melting temperature, $T_{mNiO} = 1990^0$C.

At last stage of the forming process when capacitance discharge will be finished the temperature of the hot region in oxide matrix for time $\tau_T \leq 10^{-10}$ should drop to values which would be defined by heat production due to current compliance. For estimation of this temperature we can use Eq.1 with the same problem geometry but replacing $P_{DC}$ on $P_C$, which yields $\Delta T_{DC} \leq 100^0$C.

Several processes may develop during high temperature stage of the forming but their importance will be defined by a parity of their time scales to capacitance discharge time. Evidently we should consider melting, thermoreduction of oxide, reoxidation, diffusion and solidification of the component of the reduction-oxidation reaction. Part of these processes will going in parallel and interdependently but their parity can be defined by separate consideration of temporary evolution of each process.

The melting time, $t_m$, can be evaluated from quasistationary approximation of the Stefan problem of cylindrical body melting due to a line heat source of strength $Q_C$ at $r = 0$. The appropriate solution is given by the equation [15] : $t_m = \pi a^2 \gamma_{NiO} L_{fNiO}/Q_C$, where $L_{fNiO}$ is oxide latent heat of fusion. Taking $L_f = 0.78$ kJ/g , $t_{melt} = 10^{-13}$s. We have to conclude that during high temperature forming stage the NiO conductive domain and some region around it should be transformed to melting state.

Second process which should be considered at high temperature forming stage is oxide reduction. Extensive studies of the NiO reduction have appeared in literature and the important result in frames of our consideration is that the reduction of NiO is irreversible, since the equilibrium constant $K_{eq}$ of the reduction reaction reaches $10^3$ in high temperature limit [16]. Note that oxide reduction due to direct thermal decomposition is reaction-limited process and we can neglect diffusion of the reaction products for estimation of the reduction time scale. Consider NiO reduction as first-order reaction with respect to Ni we can write solution of the reaction kinetic equation as: $C_{Ni}/C_{NiO} = [1-exp(-kt)]$, where $C_{Ni}$ is Ni concentration, $C_{0NiO}$ is initial NiO concentration, and $k = k_0 \exp(-E_{rmol}/RT)$, where $k$ is reduction reaction rate constant $E_{rmol}$ is molar activation energy and $R$ is gas constant. Using experimental dates: $E_{rmol} = 90$ kJ/mol and $k_0 = 6*10^{13}$ s$^{-1}$ [17] we can estimate the characteristic time constant of the NiO reduction as $\tau_R = 1/k < 10^{-11}$s. We have to conclude that NiO melting region and nearest solid state region with sufficiently high temperature must be converted to mixture of the Ni and O atoms during capacitance discharge regime.

The characteristic time, $\tau_D$, of the concentration gradient-driven diffusion can be written as [18] $\tau_D = l_D^2/D$ where $D$ is diffusion coefficient and $l_D$ is the characteristic distance scale. Accepting for Ni diffusion in melt NiO $D = 10^{-8}$ cm$^2$/s, and with $l_D = a = 5$ nm, $\tau_D > 10^{-5}$ s, and we can conclude that Fickian Ni diffusion and especially O diffusion, as a slower process, is not crucial for forming.

The present of the strong temperature gradients can results to temperature gradient-driven diffusion (thermomigration). Thermomigration in solid is small and therefore one is usually can be neglected as compared to concentration diffusion. In a heat flow transient induced by electrical discharge, however, temperature gradient is order to $10^8$ $^0$C/cm and thermal diffusion contribution cannot be excluded, especially in melt state of the oxide. If a homogeneous binary compound is placed in a temperature gradient, a redistribution of the constituents can occur with one constituent migrating to the cold end of the specimen and other to the hot end. This phenomenon is called Soret effect [19]. The direction of the migration and values of the mass flows is defined by the transport heat f of the diffusing ions Q$^*$. The values of the Q$^*$'s for Ni and O thermomigration in NiO are unknown. However, we can use the approaches which were developed for liquid conductive compounds [20]. Indeed, in this theory assuming that the liquid is a dense gas and applying the thermo-transport theory in binary gas mixtures the direction of the diffusion is determined primarily by the mass differences, the lighter component migrates to the warmer end and the heavy component to cold end. With this in mind, we can assume that the Ni ions migrate towards the hot region, whereas the O ions diffuse to periphery of the melt region. As a consequence, a temperature gradient drives the establishment of concentration gradients. In the stationary state this concentration gradient depends on the boundary conditions. As melt region are closed for the exchange of oxygen with the surrounding gas phase, process end up with zero atom fluxes, defining the so-called Soret state with Ni rich region in center of the melt.

The data given on Fig.3 confirm an opportunity of an establishment of the Soret state at high temperature stage of the forming. The presented dates are SIMS images of the O and Ni distribution near NiO-Pt interfaces for initial oxide structure and after forming. We can see that only O diffuses away from local nonhomogeneous regions of the NiO during forming. Assuming that these local regions have highest conductivity and, as consequence, high temperature due to Joule hitting the atoms redistribution can be defined by thermomigration and Soret state establishment.

At last forming stage, when liberation of the capacitance energy will be finished, the temperature drops to above estimated low values due thermal conductivity and the solidification of the melting region should be happened in time $t_s$. This time can be estimated from time dependence of the solidification front position $R(t)$. In our case temperature difference between solid and liquid phase near interface is not so big and we can assume that liquid has melting temperature and temperature profile in the solid is linear. The solution of the appropriate Stefan problem can be written as [14,15]: $t_s = a^2 \gamma_{Ni} L_{fNi}/2K_{Ni} T_m$. $t_s$ is less then $10^{-11}$s and fast solidification should quench the Ni filament inside oxide matrix.

The low values of the diffusion coefficient for Ni diffusion in NiO and electrode materials on last low temperature forming stage [21] allow assuming that the influence of Ni diffusion on final Ni filament size can be neglected. The oxidation on Ni-NiO interface also should be neglected because for this interface reaction at low temperature regime rate is limited by slow oxygen diffusion transport toward the NiO-Ni interface [17].

The strict solution of the problem of the Ni filament size, $R_f$, definition is based on considering of the differential form of the energy conservation equation but the simple estimations show that heating and heat transfer terms are much less in comparison with melting and chemical reaction terms. Assuming that volume of the melt is $v = \pi R_f^2 \delta$ and that intensive thermal reduction is going only in melt region we can write more simple integral energy conservation equation for steady-state regime

$$E_C = Q_{melting} + Q_{reduction} = \gamma_{NiO} v L_{fNiO} + \gamma_{NiO} v \frac{E_{Rmol}}{M_{mol}}, \quad (6)$$

where $M_{mol}$ is NiO molar mass. The solution eq.6 yields

$$R_f = \sqrt{\frac{E_C}{\pi \delta \gamma_{NiO}(L_{fNiO} + \frac{E_{Rmol}}{M_{mol}})}} \quad (7)$$

and we obtain $R_f \approx 7$ nm.

Thus we can conclude that the Ni melt filament with radius $R_f$ is formed inside NiO during energy liberation stage. After discharge the fast solidification of the Ni melt should be happened that will provide stable metallic LRS of the oxide structure after voltage torn off.

For check of model it is interesting to calculate the dimension of the metallic filament which would have the known resistance (from Fig.1 $R_{ON} \approx 50$ Ω for Ohmic section of the IVC). The temperature dependence of the LRS resistance [22] shows that Mathiesen's rule is valid and the main metal resistivity component, $\rho_{Ni}$, is defined by intrinsic Ni properties for temperatures more then 50 K. Writing total resistance as $R = \rho_{Ni} \delta / \pi R_f^2$ and taking $\rho_{Ni} = 6.9 \ast 10^{-6}$ Ohm*cm we obtained $R_f = 5$ nm and we can conclude that experimental and model filament sizes estimations well coincide.

Now consider briefly reverse transient in which the structure is passed from LRS to HRS with semiconductor conductivity but whose resistance is less on few orders in comparison with initial (before forming) state. Before this transition IVC in LRS follows Ohmic law that shows absent any barriers on interfaces or in bulk oxide. Logical base for development of LRS-HRS transient model will be the assumption of the filament rupture by the enough high current passing through structure. In the assumption that all current passes through the metal filament estimation of the current density gives value $10^9$ A/cm$^2$. Such high value of the current density will cause significant filament heating and all other possible processes will be modified by this high temperature. Let's notice, that the time constant for temperature growth remains the same as at a forming stage and has value less then $10^{-10}$s that means we can use the steady-state approximation. Using the same geometry and the same arguments for a choice of the basic direction of heat transfer that was applied for calculation of temperature at last forming stage we can use Eq.1 for temperature estimations. The calculation is shown that filament temperature before switching in HRS $T_f \approx 400^0$C and we can conclude that filament has enough high temperatures but Ni melting temperature are not achieved.

The several processes may rupture metal filament but their importance can be checked by a parity of their time scales to experimental time of the transient, which for DC biasing has view microseconds.

One of the probable processes which can break off the metal filament and return structure to HRS with semiconductor conductivity is high temperature oxidation. In situation when Ni filament is surrounded with a thick layer of oxide we can applied the Wagner model of the thermal oxidation [17]. Validity of this approach is proved by absence of the direct atmosphere - Ni interface and absence of the strong electric fields in normal to NiO-Ni interface direction. We can write the Wagner parabolic kinetic equation as $\Delta X^2 = k_p t$, where $\Delta X$ is new oxide thickness and $k_p$ is parabolic constant rate. Using the maximal value $k_p = 10^{-10}$ cm$^2$/s [17] the time for oxidation of the $5*10^{-7}$cm Ni specimen (half of the filament diameter) is $2.5*10^{-3}$ s. We can conclude that direct oxidation of Ni filament is an important process in transition from metallic to semiconductor state but it does not determine threshold conditions of the ON-OFF switching.

The instability induced by concentration or thermal gradients-driven radial diffusion is ruled out because filament temperature is low as was being shown above.

Summarized presented arguments we can conclude that any radial diffusion mass flux can not be driving force for ON-OFF instability. On the other hand, well known that the most serious and persistent reliability problem in interconnect metallization in VLSI circuits is metal atoms electromigration. The typical current density in interconnect lines of this devices achieves values $10^6$ A/cm$^2$. Such current density can cause directional mass transport in the line at the device operation temperature of 100 °C and lead to void formation at the cathode and extrusion at the anode. During ON-OFF switching in NiO with Ni filament size $R_f = 10^{-6}$ cm the current density is about $10^9$ A/cm$^2$ and electromigration may have determining significance in filament rapture process.

There is high temperature domain inside the Ni filament. In ideal situation this domain should be located in the center of the filament but really its arrangement will be adhered to filament site with the highest resistance (interfaces, geometrical constriction, compositional disordering). Taking into attention the low filament size in comparison with high temperature distribution scale we can assume that filament and the electrodes area adjoining to them has identical temperature T= $T_m$. In this case we can neglect termomigration process owing to small values as termomigration flux and flux divergences together.

In face-centered-cubic metals, such as Ni, atomic diffusion is mediated by vacancies. A flux of Ni atoms driven by electromigration to the anode requires a flux of vacancies in the opposite direction. The diffusion coefficient for Ni self diffusion and Ni diffusion in Pt is match greater then diffusion coefficient for Pt diffusion in Ni and we can neglect Pt diffusion in Ni filament [26]. In this case the vacancy flux will be stopped on cathode interface because there is not the counter atoms flux trough this boundary and vacancy will supply continuously on cathode interface. Now we come to well known diffusion problem for mass flux along the length of the one dimensional metal line due to both the electromigration driving force and Fickian diffusion [27]. The time evolution of the Ni atoms concentration with time along filament length can be obtained by solving the continuity equation, which with assuming that $D$ for Ni selfdiffusion and for Ni interdiffusion in Pt doesn't differ sufficiently, can be written as

$$\frac{\partial N_{Ni}}{\partial t} = D\frac{\partial^2 N_{Ni}}{\partial x^2} - \frac{DZ^* q\rho j}{kT}\frac{\partial N_{Ni}}{\partial x} = D\frac{\partial^2 N_{Ni}}{\partial x^2} - v\frac{\partial N_{Ni}}{\partial x} \quad (11)$$

where $Z^*$ is effective charge constant which differs from actual ion charge constant due to electron wind, $q$ elemental charge, $\rho$ resistivity of the Ni and $j$ electron current density $v$ is drift velocity.

The approximate solution of this advective-diffusion equation can be obtained in assumption that at $t = 0$ the fixed mass $M$ is released by instantaneous volume source over range $x_1 < x < x_2$, producing the initial concentration $N_{0Ni}$ given by $N_{0Ni} = M/A(x_2 - x_1)$, where $A$ is a cross-sectional area perpendicular to the $x$-axis. The solution of such a problem is readily deduced by considering the extended distribution to be composed of an infinite number of the moving instantaneous point sources and by superposing the corresponding infinite number of elementary solutions [18].

$$N(x,t) = \frac{N_{0Ni}}{2}[erf(\frac{(x-x_1)-vt}{\sqrt{4Dt}}) - erf(\frac{(x-x_2)-vt}{\sqrt{4Dt}})] \quad (13)$$

The calculations of the appropriate concentration distribution space-time evaluations have carried out with assumption that filament temperature is $400^0C$ and diffusion coefficient for this temperature is $10^{-10}$ cm$^2$/s. The values, $j$, and $\rho$ were chosen as discussed above.

The appropriate Eq. 13 concentration field space – time date, contained in Fig. 4, allows description of the filament modification as enough fast diffusion spreading of the initial Ni distribution which is conjugate with the common distribution movement in the direction of the anode under action of the electron wind.

Let's note that ON-OFF switching looks as metal-insulator transition (MIT) in part of the metal filament. The transition between metallic and thermal activated behavior of the electrical conductivity caused by the variation of chemical composition in compounds composed only from metal atoms widely discussed. It was show that MIT is impossible for system with high concentration of the delocalized electrons (pure standard metals and the majority of alloys). Nevertheless in systems with the lowered concentration of the delocalized electrons such transitions are possible [23]. Using the common approach to MIT in disordered metal (Anderson transition) with possible strong electron correlation (Mott transition) we can critical condition in Anderson and Mott models can be written in universal form as $r_H * N_C^{1/3} = a$, where $N_C$ is critical atomic concentration and $r_H$ Bohr radius and $a$ model parameter. The parameter $a$ differs for each models but at first approximation one can be chosen as 0.26 [24].

For many disordered systems the $r_H$ changes within the ranges of 0.1-1 nm and taking $r_H = 0.5$ nm we can calculate that critical atomic concentration for MIT for both transitions is equal $N_C \approx 10^{20}$cm$^{-3}$. Now we can assume that if in some part of the filament concentration will fall below this limit there will be a local transition in an insulator state. Also we should note that dimension of this region in current flow direction should be enough high for tunnel non-transparent behavior. We can assume that thickness' of this insulator domain across current flow direction should be $\delta_I > 2$ nm.

Now we have two criterions for inducing of the transition in HRS: creation of the local high resistance domain with $\delta_I > 2$ nm and $N_{Ni} < N_C$. Then taking into attention that ON-OFF switching time for DC regime of the switching has values within the several microseconds, the above criterions should be connected with microsecond time scale for potential processes of the filament rupture.

For detail analysis of the time evolution concentration instability near cathode we can take of $x_1 = \infty$ and $x_2 = 0$, the graphs of the appropriate solutions of the Eq. 12, which are shown on Fig. 5, demonstrate that the MIT criterion for layer with few nanometers thickness is achieved for microsecond scale times and therefore the electromigration and Fickian diffusion is the most significant processes bringing of the metallic conductivity destruction.

Let's emphasize that this process has features of the mechanism with positive feedback - the formation of the region with lowered metal atom density even without MIT should increase local filament resistance that leads to local temperature increase and to local diffusion and electromigration acceleration. We have mechanism of the instability with voltage-controlled NDR. Also it should be note that if thin layer with nonmetallic properties was formed, the domain with high electric field will be built in filament structure. This part of the Ni filament will be quickly transformed in NiO by the fast electric field enhanced thermal oxidation [17].

The final problem to be discussed is OFF-ON transient. The phenomenological pictures of the forming and HRS-LRS transition coincide, that allows assumption of the generality of the mechanisms of the both phenomena. As well as forming, the OFF-ON transition can be classified as hard breakdown of the insulator NiO layer which was formed near cathode interface during ON-OFF switching. Also we can assume that initial breakdown mechanism is not so important for restoration of the Ni filament and main processes should be developed on second stage of the breakdown.

The estimations energy which was storage in capacitor structure before breakdown with the same assumptions as for forming gives $E_C = 10^{-14}$ J and with average capacitance discharge time $\tau_0 = 10^{-9}$ s the density of the power $P_{DC} = 10^{15}$ W*cm$^{-3}$.

Here we should change the problem geometry because cylindrical approximation of conducting cord with primary heat transfer to a radial direction will not be valid already. A correct estimation of the processes times and temperature values can be expressed by considering that energy dissipation occurs in spherical oxide region witch radius is $\delta_i$. The thermal constants of the NiO, Ni, and Pt differ among themselves within the limits of one order of magnitude and we have considered problem of the hot NiO sphere surrounded with a thick cold NiO layer. Calculations have shown that main processes (heating, melting, NiO reduction) has the same time scales as for forming and it is allows to estimate size of the new metal region, $R_f^*$, with using of the simple integral energy balance equation as Eq. 7 but with $v = 4\pi R_f^3/3$. The $R_f$ is thus

$$R_f = \sqrt[3]{\frac{3Q_C}{4\pi\gamma_L(L_f + \frac{E_{Rmol}}{M_{mol}})}} \quad (15)$$

Eq. (13) yields $R_f = 8$ nm and we can conclude that after OFF-ON transition the Ni filament would be practically completely restored.

In this paper we have developed a model of the growth and rupture nanosized metal filament inside oxide matrix. The dielectric breakdown of the oxide with conditions of the adequate compliance current and capacitance discharge of the energy, which have been stored in thin film oxide capacitor structure, initiates the sharp local temperature growth. The analysis of the characteristic time scales of the possible high temperature enhancement processes have shown that metal cluster can be formed due to oxide reduction and temperature gradient-driven diffusion. The Soret state with metal segregation on the center of the high temperature region is quenched by fast temperature drop after capacitance discharge ending. The next biasing induces fast metal atom electromigration due to high electron current density. The metal concentration near cathode region is depleted, and, after achievement of the metal-insulator transition criterion, the tunnel non-transparent local domain with high resistance and high electric field is arisen. This part of metal filament is transformed in oxide by the subsequent fast electric field enhanced thermal oxidation.

The presented model can explain nonvolatile resistance switching of the oxide thin films structure which is promising phenomenon for development of the high density stackable nonvolatile memory.

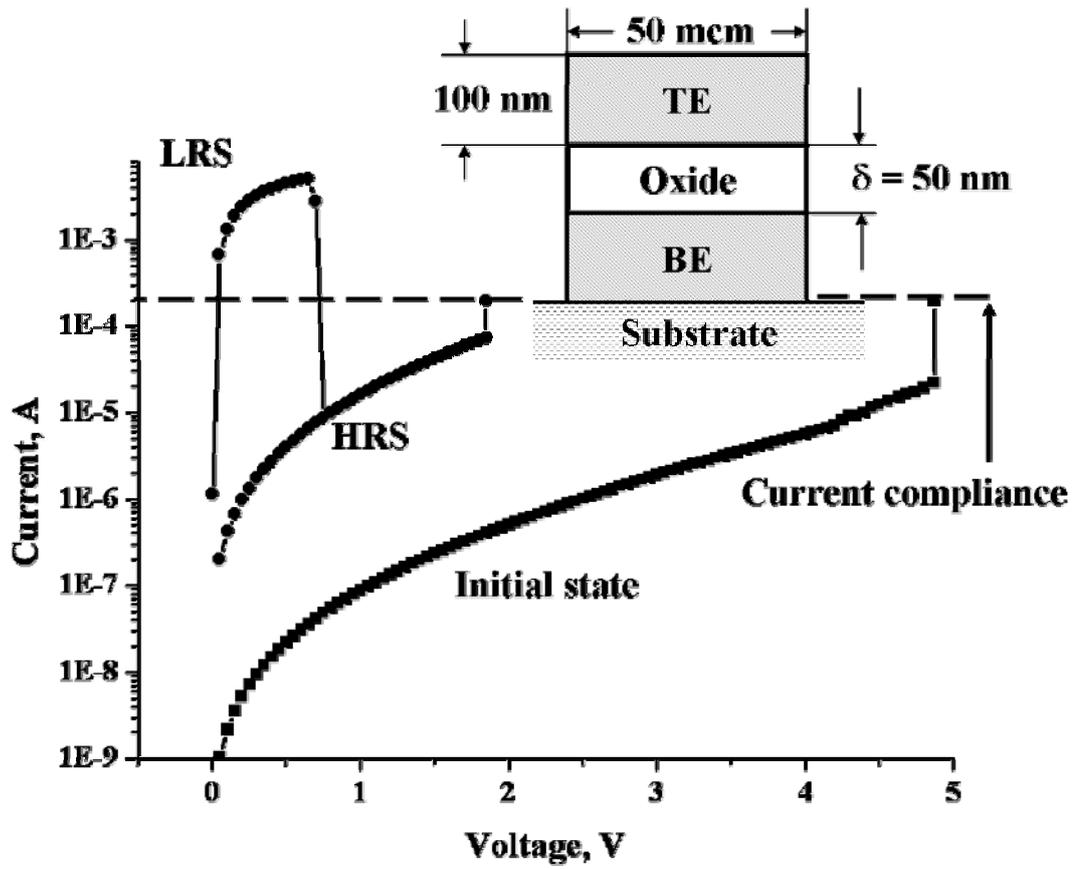

Figure 1. Typical current-voltage characteristic for Pt-NiO-Pt structure with nonvolatile unipolar switching in voltage controlled regime of the measurement. TE and BE are Pt top and bottom electrodes.

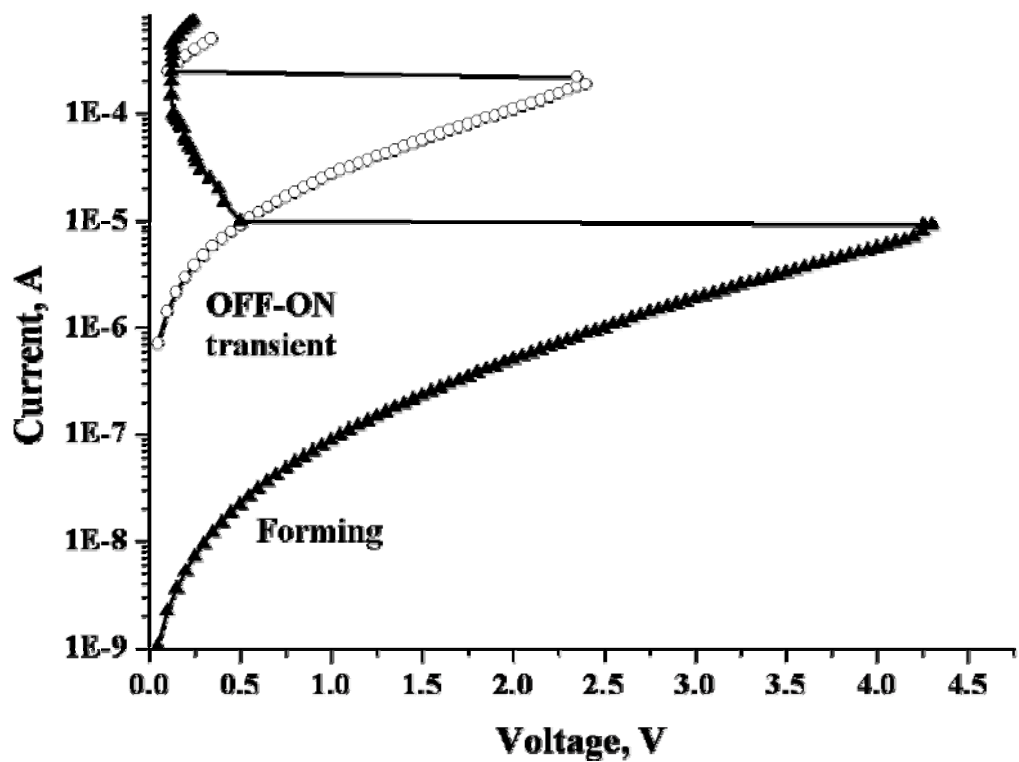

Figure 2. Current-voltage characteristic for forming and OFF-ON transient in current controlled regime of the measurement.

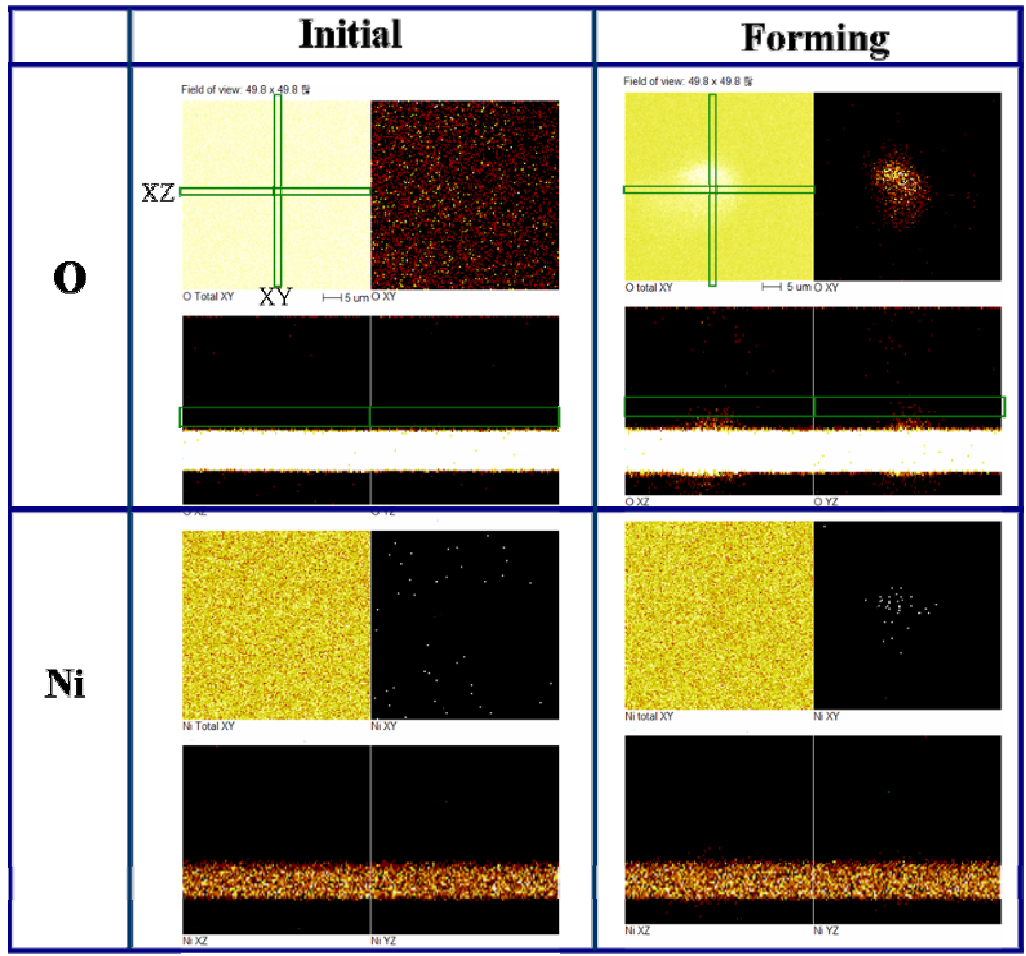

Figure 3. SIMS images of the Ni and O distributions near NiO-Pt interfaces in initial state and after forming

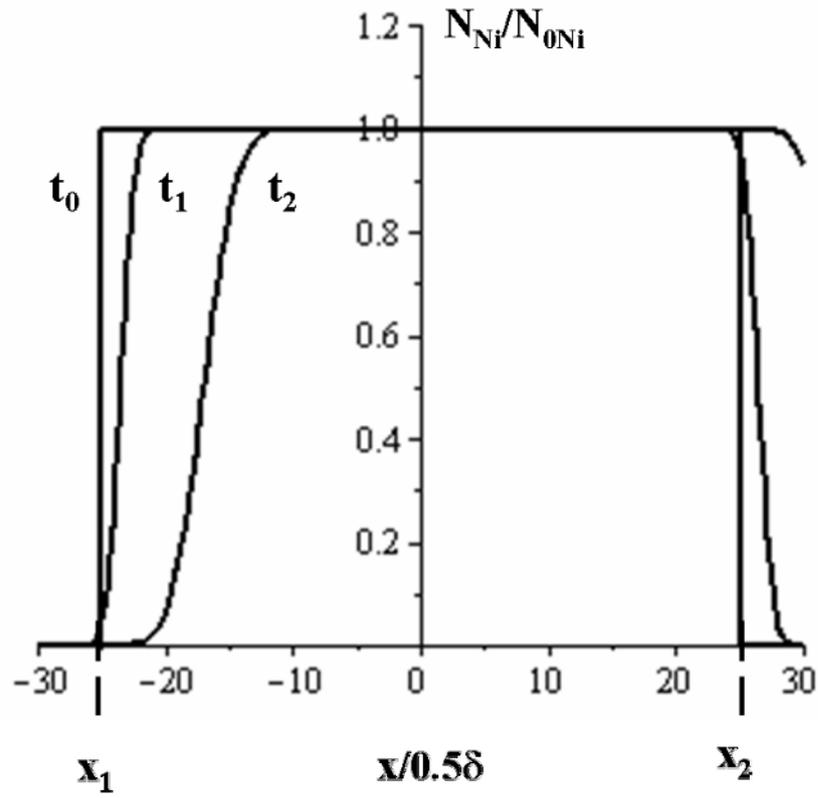

Figure 4. The plots of eq.(13) for selected times. $t_0=0$ (initial distribution), $t_1=8*10^{-7}$s, $t_2=2.8*10^{-6}$s.

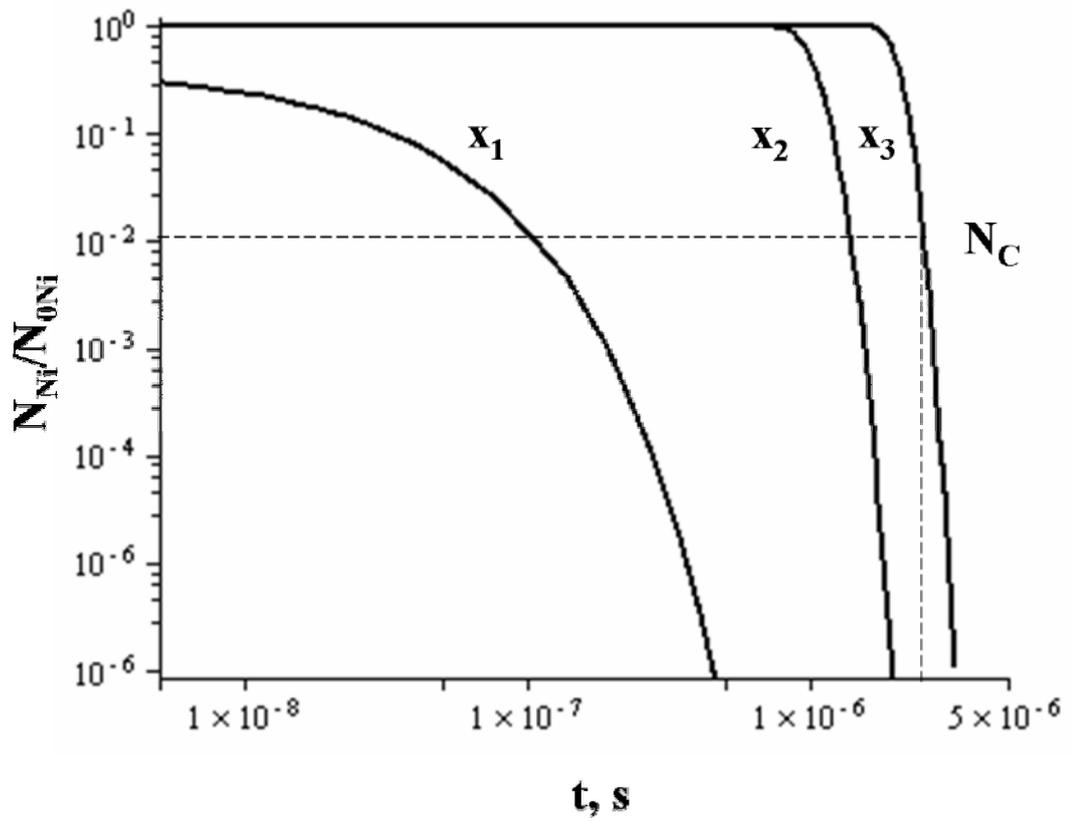

Figure 5. The plots of the eq.14 for selected $x$. $x_1 = 0$; $x_2 = 10^{-7}$ cm; $x_3 = 2*10^{-7}$ cm;

: